\documentclass[10pt]{artikel3}
\usepackage{amsfonts,amsmath,amssymb,hyperref}

\usepackage[dvips]{feynmp}

\usepackage{epsfig}

\usepackage{float,braket}
\usepackage[adobe-utopia]{mathdesign}
\usepackage[T1]{fontenc}
\usepackage{subfigure}

\usepackage{graphicx,wrapfig,color}

\newcommand{\p}{\partial}

\renewcommand{\d}{\ensuremath{\mathrm{d}}}

\renewcommand{\d}{\ensuremath{\mathrm{d}}}

\definecolor{Rood}{rgb}{1, 0, 0} 

\begin{document}
\title{\noindent {\bf Pad\'{e} approximation and glueball mass estimates in $3d$ and $4d$ with $N_c=2, 3$ colors}}
\author{D.~Dudal$^{a}$\thanks{david.dudal@ugent.be},\; M.S.~Guimaraes$^{b}$\thanks{msguimaraes@uerj.br},\; S.P.~Sorella$^{b}$\thanks{sorella@uerj.br}\\\\
{\small  \textnormal{$^{a}$ Ghent University, Department of Physics and Astronomy, Krijgslaan 281-S9, 9000 Gent, Belgium}}
\\
\small \textnormal{$^{b}$ Departamento de F\'{\i }sica Te\'{o}rica, Instituto de F\'{\i }sica, UERJ - Universidade do Estado do Rio de Janeiro}
 \normalsize}

\date{}
\maketitle
\begin{abstract}
A Pad\'{e} approximation approach, rooted in an infrared moment technique, is employed to provide mass estimates for various glueball states in pure gauge theories. The main input in this analysis are theoretically well-motivated fits to lattice gluon propagator data, which are by now available for both $SU(2)$ and $SU(3)$ in 3 and 4 space-time dimensions. We construct
appropriate gauge invariant and Lorentz covariant operators in the (pseudo)scalar and (pseudo)tensor sector. Our estimates compare reasonably well with a variety of lattice sources directly aimed at extracting glueball masses.
\end{abstract}

\setcounter{page}{1}

\section{Introduction}
Although confinement is a well accepted phenomenon in pure gauge theories \cite{Greensite:2011zz}, the extraction of the observable degrees of freedom, which ought to be glueballs, is a challenging task. Several theoretical methods\footnote{Sometimes with the inclusion of quarks to study the glueball spectrum in QCD itself} have been tested, to name a few: qualitative studies \cite{Jaffe:1985qp}, effective Hamiltonian methods \cite{Buisseret:2009yv,Buisseret:2013ch,Szczepaniak:2003mr,Kaidalov:1999de}, AdS/CFT inspired tools \cite{Brower:2000rp,Colangelo:2007pt}, lattice simulations \cite{Teper:1998kw,Morningstar:1999rf,Chen:2005mg,Gregory:2012hu}, functional approaches \cite{Strauss:2012kqa,Meyers:2012ka}, Regge trajectory analyses \cite{LlanesEstrada:2000jw}, sum rules analyses \cite{Narison:1996fm}, etc. We refer to \cite{Mathieu:2008me} for a recent review on the subject. Also, from the  experimental side, the status of glueballs is at the best inconclusive as they are hard to detect, partially due to their mixing with other states, see \cite{Crede:2008vw} for more details.

In this current note, we will apply a method developed in \cite{Dudal:2010cd}. The main purpose is to benefit from high precision lattice computations of the gluon propagator in certain preferential gauges, in particular the Landau gauge \cite{lattice}. As these correlation functions carry essential nonperturbative information on the gluon dynamics, it seems natural to benefit from these data. As we are interested in continuum computations, we need functional forms for e.g.~the gluon propagator, as we plan to study the correlation functions of bound state and thus of composite operators. As it  will become clear, we will probe the analyticity properties of the latter propagators, which is not an easy task if the input gluon propagator is a complicated function. A recent numerical approach to derive the spectral density of a (tree level) bound state propagator given an a priori analytical prescription for the input constituent propagator can be found in \cite{Windisch:2012sz,Windisch:2012zd}, thereby confirming the analytical results of \cite{Dudal:2010wn}. A far more appealing approach would be to only use the gluon lattice data, which is however still in its infancy given all the  difficulties to extend the lattice data from the Euclidean region $p^2\geq 0$ to the complex $p^2$ plane \cite{Dudal:2013vf}.

We will thus rely on the fits constructed for $d=4$ $SU(3)$ data in \cite{Oliveira:2012eh} (see also \cite{Dudal:2010tf,Dudal:2012zx}) and for $d=3,4$ $SU(2)$ data in \cite{Cucchieri:2011ig}, which have the upshot of allowing for a pure analytical study of the correlation functions over the complex plane. It is worth pointing out that these fits are the result of a well-motivated and consistent theoretical framework. We remind here that the Landau gauge, as any covariant gauge fixing, suffers from the Gribov problem: there exist multiple gauge equivalent field configurations fulfilling the  same gauge condition. An effective action formalism to deal with the Gribov issue  was worked out in a series of papers by Gribov \& Zwanziger \cite{Gribov:1977wm,Zwanziger:1989mf,Zwanziger:1992qr,Zwanziger:1993dh}, see also the recent work \cite{Pereira:2013aza}. Basically, the domain of integration of the gauge fields in the Euclidean functional integral is further constrained to the first Gribov region $\Omega$, whose boundary is the Gribov horizon, where the Faddeev-Popov operator attains the first vanishing eigenvalue \cite{Vandersickel:2012tz}. In recent years, we included into the original derivation of \cite{Gribov:1977wm,Zwanziger:1989mf,Zwanziger:1992qr,Zwanziger:1993dh}, the dynamical effects of dimension two condensates \cite{Dudal:2007cw,Dudal:2008sp,Dudal:2008rm,Dudal:2011gd}, resulting in what is nowadays called the Refined Gribov-Zwanziger (RGZ) action.

Let us also notice that the relevance of dimension two condensates for certain nonperturbative effects in gauge theories was already realized in e.g.~\cite{Boucaud:2001st,Gubarev:2000eu,Gubarev:2000nz,Verschelde:2001ia,Kondo:2001nq}. The analytic form of the tree level gluon propagator obtained in the Refined Gribov-Zwanziger theory is, in general, given by
\begin{equation}
\mathcal{D}(p^2) = \frac{p^4 + 2 M^2 p^2+ M^4 - \rho \rho^\dagger}{
                    p^6 + p^4 \left(m^2 + 2 M^2 \right) + p^2 \left(2 m^2 M^2 + M^4 + \lambda^4 - \rho \rho^\dagger \right)
                  + m^2 \left( M^4 - \rho \rho^\dagger \right)
                  + M^2 \lambda^4 - \frac{\lambda^4}{2}
                  \left( \rho + \rho^\dagger \right)} \; ,
\label{Dgl}
\end{equation}
where $m^2$, resp.~$M^2$ and $\rho$ (with c.c.~$\rho^\dagger$) are corresponding to the $d=2$ gluon condensate $\braket{A^2}$, resp~other $d=2$ condensates constructed from the additional fields present in the Gribov-Zwanziger action, let us refer the interested reader to the paper \cite{Dudal:2011gd} for details. The quantity $\lambda^4$ is directly related to the restriction to the Gribov region $\Omega$. If it happens that $\rho$ is real, then the RGZ propagator can be rewritten into the form
\begin{equation}
\mathcal{D}(p^2)=\frac{p^2+M_1^2}{p^4+M_2^2p^2+M_3^4}\;.\label{Dg2}
\end{equation}
The $d=2$ mass parameters $M_1^2$, $M_2^2$ and $M_3^2$ are recombinations of $m^2$, $M^2$, $\rho$ and $\lambda^4$, whereby $M^2$ and $\rho$ appear in a fixed combination \cite{Dudal:2011gd}. We feel it is important to point out here that the form of the propagator \eqref{Dgl} is in general predicted by the RGZ formalism. The mass parameter $\lambda^4$ is present because of the restriction to the Gribov region $\Omega$, the others are related to a stabilization of the vacuum by means of $d=2$ condensates. Thus this tree level analytic form is dictated by the underlying RGZ dynamics. In principle, these vacuum expectation values are determined by self-consistent gap equations, that is by minimization of the effective potential (vacuum energy). Such program was carried out first in \cite{Dudal:2008sp,Dudal:2011gd}, albeit in a one-loop approximation, giving a qualitative but not always a superb quantitative agreement with the data. This is acceptable, since it is evidently not a straightforward task to compute the effective potential to arbitrary high order. The key features of the formalism are clear at lowest order already. Though, for the current purposes we also require a quantitative estimate for the condensates, as these are exactly the mass parameters that will fuel our eventual glueball mass estimates. We will therefore make use of the gluon lattice data to attribute values to the condensates, in particular have the functions \eqref{Dgl} and \eqref{Dg2} been used as fitting proposals in the papers \cite{Oliveira:2012eh,Cucchieri:2011ig} on which we shall thus rely.

It is worthwhile to remember that other functional approaches exist for the study of Yang-Mills Green functions in the Landau gauge, in particular the Schwinger-Dyson equations that corresponds to the quantum equations of motion. Also these equations are impossible to solve exactly, but can be replaced after certain approximations by (numerically) solvable equations. These numerical propagators can then also be \emph{fitted} with functions as in \cite{Alkofer:2003jj,Aguilar:2011ux}. One can even directly attempt to construct numerical estimates for the gluon spectral function based on either the Schwinger-Dyson equations \cite{Strauss:2012dg} or ``inversion'' of the lattice data \cite{Dudal:2013yva}, but in none of the aforementioned cases a closed analytical expression can be derived, one is always reduced to fitting the numerical result with some a priori completely free to choose function. This is different from the RGZ (loop expansion) approach, where the functional form are closed analytical expressions (albeit with the condensates' values fixed via the lattice data).

In Sect.~2, we will first describe which composite operators $\mathcal{O}$ we need to describe specific glueball states and we will construct the spectral density of the corresponding two-point correlation function $\braket{\mathcal{O}(p)\mathcal{O}(-p)}$ in a first order approximation, also known as the Born approximation \cite{Windisch:2012sz}. We furthermore list the values, as reported and discussed in other works, for the RGZ parameters appearing in either \eqref{Dgl} or \eqref{Dg2}. In Sect.~3, we spend a few words on the infrared moments technique and how it can be used to get a first rough estimate of glueball masses. We end with a discussion in Sect.~4.

\section{The gluon propagator input, the glueball operators and associated spectral densities}
\subsection{The RGZ gluon propagator}
Let us first give some numbers we will need later on. For the $SU(3)$ case, we rely on the RGZ fitting parameters of \cite{Oliveira:2012eh}\footnote{Extrapolated to infinite volume with $\beta=6.2$.}, where the expression \eqref{Dg2} seems to be singled out by the data\footnote{We will always omit the necessary global renormalization factor as it will play no role in the current paper.}:
\begin{equation}\label{a-1}
     M_1^2= 4.473(21)~\text{GeV}^2\,,\qquad M_2^2= 0.704(29)~\text{GeV}^2\,,\qquad M_3^4= 0.3959(54)~\text{GeV}^4\;.
\end{equation}
With these values, the propagator \eqref{Dg2} displays two c.c.~poles, namely at
\begin{equation}\label{a0}
    -p^2=\mu^2\pm i\sqrt 2\theta^2\approx 0.352\pm i0.513
\end{equation}
in appropriate GeV units. To obtain the location of the poles, we always use the central value of the fitting estimates as in \eqref{a-1}. The one standard deviation errors on those are taken from the original papers and shown between parentheses.

This observation of c.c.~poles is what lies at the heart of the $i$-particles setup introduced in \cite{Baulieu:2009ha}: the RGZ gluon propagator can be expressed in terms of a pair of ``complex'' particles with c.c.~masses. Clearly, such degrees of freedom are not physical, hence there is no direct observable information in the Landau gauge gluon propagator.

As discussed in \cite{Cucchieri:2011ig} for $SU(2)$ lattice data\footnote{Obtained with  $V=128^4$ as the largest volume and  with $\beta= 2.2$.}, the corresponding fitting parameters are given by
\begin{equation}\label{a-1b}
     M_1^2 = 2.508(0.070)~\text{GeV}^2\,,\qquad M_2^2= (0.768(0.017))^2~\text{GeV}^2\,,\qquad M_3^4= (0.720(0.009))^2~\text{GeV}^4\;,
\end{equation}
leading to the roots (GeV units) \cite[Table IV]{Cucchieri:2011ig}
\begin{equation}\label{a0}
    -p^2=\mu^2\pm i\sqrt 2\theta^2\approx 0.29\pm i0.66\;.
\end{equation}
For $SU(2)$, there are also data available in $d=3$. Interestingly, there, the full RGZ propagator \eqref{Dg2} is needed to adequately describe the lattice data. For convenience, we write it in the following form (GeV units)
\begin{eqnarray}\label{a2c}
\mathcal{D}(p^2)&=&\frac{\alpha}{p^2+\omega_1^2}+ \frac{\beta}{p^2+\omega_2^2}+\frac{\beta^\dagger}{p^2+\omega_2^{2\dagger}}\,,\\&& \alpha=-0.024(5)\,,\omega_1^2= 0.046(4)\,, \beta= 0.216(3)+i0.27(5)\,,\omega_2^2= 0.215(5)+i0.580(6) \equiv \mu^2+i\sqrt{2}\theta^2\;.\nonumber
\end{eqnarray}
The $d=3$ gluon propagator thus displays next to two  c.c.~roots also a relatively small real root. Though, since its residue is negative, it neither describes a physical degree of freedom.

\subsection{(Pseudo)scalar and (pseudo)tensor glueball operator in $d=4$}
As we wish to work in the context of local Lorentz-invariant\footnote{In practice, we employ an Euclidean space-time.} quantum field theory, we are looking for gauge invariant/Lorentz covariant operators, constructed in such a way that they described states with specific $J^{PC}$ quantum numbers. This occasionally necessitates the introduction of projection operators onto the desired subspace. For the scalar operator ($J^{PC}=0^{++}$), we can take $\mathcal{G}^{0++}=F_{\mu\nu}^2$, while for the pseudoscalar state ($J^{PC}=0^{-+}$) the corresponding operator is given by  $\mathcal{G}^{0-+}=F_{\mu\nu}\widetilde{F}_{\mu\nu}\equiv \frac{1}{2}\varepsilon^{\mu\nu\alpha\beta}F_{\mu\nu}F_{\alpha\beta}$.

In the (pseudo)tensor sector, a little more effort is required. As the equations of motion derived from the (RGZ) action are somewhat different from those of the usual Yang-Mills action \cite{Zwanziger:1992qr,Dudal:2008sp}, the standard energy momentum tensor, $t_{\mu\nu}=F_{\mu\alpha}F_{\alpha\nu}-\frac{\delta_{\mu\nu}}{d}F_{\alpha\beta}^2$ does not qualify anymore as it is not necessarily conserved\footnote{This does not mean the (RGZ) theory has no conserved energy-momentum tensor, it simply differs from $t_{\mu\nu}$.}. In \cite{Dudal:2010cd,Capri:2010pg}, we proposed
\begin{equation}\label{tensor}
  \mathcal{G}^{2++}_{\mu\nu}= \p^4 t_{\mu\nu}-\p^2 \p_\mu\p_\alpha t_{\alpha\nu}-\p^2\p_\nu\p_\alpha t_{\alpha\mu}+ P_{\mu\nu}\p_\alpha\p_\beta t_{\alpha\beta}
\end{equation}
for the tensor ($J^{PC}=2^{++}$) glueball; where $P_{\mu\nu}=\p^2\delta_{\mu\nu}-\p_\mu \p_\nu$ is the transverse projector. The foregoing operator $\mathcal{G}^{2++}_{\mu\nu}$ is, irrespective of the equations of motion, conserved, symmetric and traceless and has the upshot to be directly proportional to $t_{\mu\nu}$ if it happens that $\p_\mu t_{\mu\nu}=0$.

The symmetric and conserved operator
\begin{equation}\label{2}
  q_{\kappa \mu}=\p_\lambda \p_\nu\left(F_{\kappa\lambda}F^\ast_{\mu\nu}+F_{\kappa\lambda}^\ast F_{\mu\nu}\right)
\end{equation}
can be easily made traceless, without compromising its symmetry properties as well as its conservation, namely by passing to
\begin{equation}\label{3}
  \mathcal{G}^{2-+}_{\mu\nu}=\p^2 q_{\mu\nu}- \frac{1}{3}P_{\mu\nu} q_{\alpha\alpha}\;,
\end{equation}
an operator suited to describe the $J^{PC}=2^{-+}$ state.

\subsection{Scalar and tensor glueball operator in $d=3$}
In $d=3$, the dual of $F_{\mu\nu}$ is an axial vector. We have not been able to write down in a Lorentz covariant notation a simple local operator that would describe the analogue of pseudoscalar/-tensor as in $d=4$. Therefore, for the present  work, we will limit ourselves to the scalar/tensor case where the operators $\mathcal{G}^{0++}$ and $\mathcal{G}^{2++}_{\mu\nu}$ can be immediately employed.

\subsection{Derivation of the spectral densities at lowest order}
Using the $i$-particle representation of the RGZ propagator \cite{Dudal:2010cd}, the necessary spectral representations of the afore described glueball correlators can be derived using the tools of \cite{Dudal:2010wn}. We recall here that a physical particle propagator, e.g.~$\braket{\mathcal{G}^{0++}(p)\mathcal{G}^{0++}(-p)}\equiv \mathcal{F}(p^2)$, should be consistent with a K\"{a}ll\'{e}n-Lehmann integral form, which in Euclidean conventions reads
\begin{equation}\label{KL1}
  \mathcal{F}(p^2)=\int_{\tau_0}^{+\infty}\d t\frac{\rho(t)}{t+p^2}\;,
\end{equation}
with $\left.\rho(t)\right|_{t\geq \tau_0}\geq 0$. Making $p^2$ a complex variable, eq.~\eqref{KL1} describes an everywhere analytic function, with the exception of a branch cut along the negative real axis. Using Cauchy's basic theorem, the density $\rho(t)$ is proportional to the discontinuity along the axis, which due to the optical theorem means that it must be positive, at least when we are talking about physical observables.  E.g.~for a confined gluon this does not need to be the case.

Now, given that the $i$-particles come in complex pairs and that we shall consider the single bubble approximation\footnote{All considered operators have quadratic pieces in the gluon field, hence the lowest order contribution to the considered bound state propagators will always be a bubble diagram.} to $\mathcal{F}(p^2)$, we need to carefully consider which diagrammatic subsets of the full correlator can correspond to a real degree of freedom, in the current approximation at least. As studied into great detail in \cite{Baulieu:2009ha}, if the 2 internal propagator lines of the bubble corresponds to a pair of $i$-particles with c.c.~masses, and only then, the resulting contribution is consistent with the representation \eqref{KL1}. A case by case check is then all that is needed to verify the positivity of the density $\rho(t)$. For the moment, a full-fledged approach that can consistently remove the unphysical pieces (from combining particle propagators not containing c.c.~mass pairs for example) order by order is not yet available. We will adopt the working assumption that for now, we can stick with only retaining the physical contributions to the respective correlators. Recalling that, upon considering particles with masses squared $m^2\geq0$ and $M^2\geq0$, their one-loop composite two-point correlation function will develop a branch point at $p^2=-(m+M)^2$, see e.g.~\cite{zuber}, it is clear that complex valued branch points are to be expected when propagator with complex masses are combined. This does not occur when $m^2$ and $M^2$ are c.c.~, as discussed in \cite{Dudal:2010wn,Baulieu:2009ha}.

So, using the complex mass Cutkosky rules (see \cite{Dudal:2010wn} for a discussion), we can derive for each of the above operators the physical piece of the glueball correlators. Notice that in $d=3$, also the 2 Yukawa propagators with the wrong sign, cf.~\eqref{a2c}, will contribute to the physical piece of the bound state correlator, since the 2 minus signs will combine into a physical, positive signed, piece. We will refrain from writing down the tedious calculations here, but immediately list the final spectral densities we will continue to work with.

In $d=4$, we found, up to irrelevant global prefactors,
\begin{eqnarray}
  \!\!\! \!\! \rho^{0++}_{d=4}(t)&=&\sqrt{1-8\frac{\theta^4}{t^2}-4\frac{\mu^2}{t}}\left(\frac{t^2}{2}+2\theta^4-2t\mu^2+3\mu^4\right)\mathcal{H}(t-\tau_0)\;,\label{a01}\\
  \!\!\!\!\!   \rho^{0-+}_{d=4}(t)&=&\sqrt{1-8\frac{\theta^4}{t^2}-4\frac{\mu^2}{t}}\left(2\theta^4+\mu^2t-\frac{1}{4}t^2\right)\mathcal{H}(t-\tau_0)\;, \label{a02}\\
 \!\!\!\!\!    \rho^{2++}_{d=4}(t)&=&\sqrt{1-8\frac{\theta^4}{t^2}-4\frac{\mu^2}{t}}t^2\left(16\theta^8t^2-4\theta^4\mu^2t^3+16\theta^4t^4+9\mu^4t^4-\frac{9}{2}\mu^2t^5+\frac{3}{2}t^6\right)\mathcal{H}(t-\tau_0)\;,\label{a03}\\
 \!\!\! \!\!   \rho^{2-+}_{d=4}(t)&=&\sqrt{1-8\frac{\theta^4}{t^2}-4\frac{\mu^2}{t}}t^2\left(t^2+48\theta^4-6t\mu^2\right)\left(t^2-8\theta^4-4t\mu^2\right)\mathcal{H}(t-\tau_0)\;.\label{a1}
\end{eqnarray}
The Heaviside step function $\mathcal{H}(x)$ implements the threshold which is in all cases given by the expression $\tau_0=2\left(\mu^2+\sqrt{\mu^4+2\theta^4}\right)$. It remains to check under which conditions the above spectral functions are positive. For \eqref{a01} and \eqref{a02}, it is easily checked that there are no real roots for $t>\tau_0$ and that the functions are positive for that interval given that always $\theta^2\geq0$, $\mu^2\geq0$. In the case of \eqref{a1}, the only real root for $t>\tau_0$ is given by $t=3\mu^2+\sqrt{3}\sqrt{3\mu^4-16\theta^4}$, at least when $3\mu^4\geq 16\theta^4$. It is then easily checked that this root is located below $\tau_0$ if $\mu^4\leq \frac{98}{15}\theta^4$. If the latter condition is fulfilled, then \eqref{a1} remains positive over the whole of the relevant domain. Using the numbers quoted in eq.~\eqref{a-1} for $SU(3)$ and eq.~\eqref{a-1b} for $SU(2)$, the foregoing conditions are clearly fulfilled. For the tensor case \eqref{a03}, we had to check numerically that the spectral density is positive for $t> \tau_0$.

In $d=3$, we found, up to irrelevant global prefactors,
\begin{eqnarray}
    \rho^{0++}_{d=3}(t)&=& \frac{\alpha^2}{\sqrt t}\left(8\omega_1^4-4\omega_1^2t+t^2\right)\mathcal{H}(t-4\omega_1^2) + 2\frac{\beta\beta^\dagger}{\sqrt t}\left(t^2+8\theta^4-4t\mu^2+8\mu^4\right)\mathcal{H}(t-\tau_0)\;,\label{b01}\\
    \rho^{2++}_{d=3}(t)&=&\frac{\alpha^2}{\sqrt t}\left(\frac{t^4}{8}+(4\omega_1^2+t)^2\right)\mathcal{H}(t-4\omega_1^2)\nonumber\\ && + 2\frac{\beta\beta^\dagger}{\sqrt t}\left(\frac{t^4}{8}(4\mu^2+t)^2+2t^3\theta^4(7t-4\mu^2)+8t^2\theta^8\right)\mathcal{H}(t-\tau_0)\;.\label{b02}
\end{eqnarray}
In both cases, no real roots are to be reported over the $t$-domain of interest for $\mu^2\geq0$, $\theta^2\geq0$, hence the $d=3$ spectral densities are found positive.

\section{Infrared moments, Pad\'{e} approximation theory and mass estimation}
\subsection{Setup of the moment problem}
So far, we have derived the spectral densities of a set of glueball correlators at lowest (one bubble) order. Clearly, none of the functions \eqref{a01}, \eqref{a02}, \eqref{a03}, \eqref{a1} or \eqref{b01}, \eqref{b02} displays a pole on the negative real axis, something that would correspond to a massive physical particle. This is no surprise given the approximation, usually a dynamical pole will only emerge if some kind of resummation is performed. Though, given the complexity of the gluon interactions, to our knowledge,  a self-consistent approximation scheme that would allow to resum to some extent higher order bubble graphs, let stand alone construct the spectral properties of such graphs\footnote{For the quark dynamics, a popular approximation scheme is that of Nambu-Jona-Lasinio-like (NJL) interactions based on integrating out the gluons, which allows resummation techniques in at least a large $N$ approach \cite{Buballa:2003qv}. Alternatively, one can study the Bethe-Salpeter equations for a bound state, something which has received widespread attention in the meson/baryon sector \cite{Roberts:1994dr,Alkofer:2000wg,Bashir:2012fs}. Nevertheless, similar approaches in the glueball sector are still hard to handle \cite{Strauss:2012kqa}.}, has not yet been achieved so far.

Therefore, we will adopt a different strategy here, based on a suitable moment problem. Moment problems are not new in particle physics, see \cite{Narison:1996fm} for similar approaches using the OPE/spectral methods/sum rules. The difference with what we will explain is that our moments will be IR based, in contrast with ruling approaches. Let us first provide a short survey of the IR moment problem as originally introduced in \cite{Dudal:2010wn}. We reconsider $\mathcal{F}(p^2)$ as given in eq.~\eqref{KL1}, and perform the substitution $t=\frac{1}{s}$, so that
\begin{equation}\label{b3}
    \mathcal{F}(p^2)=\int_{0}^{1/\tau_0} \frac{\rho(1/s)}{s}\frac{1}{1 +s p^2}\d s\equiv \int_{0}^{s_0}\frac{\sigma(s)}{1+sp^2}\d s\;,
\end{equation}
this expression can be easily expanded around $p^2=0$,
\begin{equation}\label{b4}
    \mathcal{F}(p^2)=\sum_{n=0}^{\infty}\int_{0}^{s_0} s^n \sigma(s)\d s (-1)^n (p^2)^n \equiv \sum_n \nu_n (-1)^n(p^2)^n
\end{equation}
where we defined the IR moments
\begin{equation}\label{b5}
    \nu_n= \int_{0}^{s_0} s^n \sigma(s)\d s\;.
\end{equation}
Notice that \eqref{b4} defines a formal power series, so it does not need to converge.

By passing to
\begin{equation}\label{b5bis}
    f(z)=\frac{1}{z}\mathcal{F}\left(-\frac{1}{z}\right)=\int_0^{s_0}\frac{\sigma(s)}{z-s}\d s\;,
\end{equation}
we arrive at a system with finite boundaries.

Before going any further, there is a technical issue to deal with. We notice that eq.~\eqref{KL1}, or equivalently eq.~\eqref{b3}, is a divergent integral, simply visible on dimensional ground through power counting. This is a typical feature of quantum field theory and the standard way to obtain a finite spectral integral is to subtract the first few orders of its Taylor expansion \cite{Narison:2002pw}. If we need $r$ subtractions at $p^2=T$, with $T$ the subtraction scale, the subtracted spectral representation reads:
\begin{equation}\label{b8}
\mathcal{F}^{\text{sub}}(p^2)\equiv (-1)^r(p^2-T)^r \int_{0}^{s_0}\frac{1}{(1+sT)^r}\frac{s^r\sigma(s)}{1+sp^2}\d s\;.
\end{equation}
We will thus consider the moment problem associated to the moments $\nu'_n$, obtainable via
\begin{equation}\label{b8b}
  \hat{\mathcal{F}}(p^2) = -\frac{\mathcal{F}^\text{sub}(p^2)}{(-1)^r(p^2-T)^r}=\int_{0}^{s_0}\frac{\sigma'(s)}{1+sp^2}\d s\,,\qquad \text{with }\sigma'(s)=\frac{s^r}{(1+sT)^r}\sigma(s)\geq 0\;.
\end{equation}
Analogously as before, we introduce
\begin{equation}\label{b8c}
  f(z)=\frac{1}{z}\hat{\mathcal{F}}\left(-\frac{1}{z}\right) = \int_{0}^{s_0}\frac{\sigma'(s)}{z-s}\d s\;,
\end{equation}
with
\begin{equation}\label{b8d}
\nu'_n= \int_0^{s_0} s^n \sigma'(s)\d s<\infty\,,\qquad \forall n\in\mathbb{N}\;.
\end{equation}
The final question is, given a set of moments $\nu'_0,\ldots \nu'_{2N-1}$, how to construct the underlying spectral density, with hopefully good convergence properties in terms of the numbers of input moments? For this so-called reduced Hausdorff problem, a nice answer has been provided in terms of Pad\'{e} approximants of order $[N,N-1]$,
\begin{equation}\label{b5tris}
    f(z)= \frac{P_{N-1}(z)}{Q_N (z)} + \mathcal{O}\left(z^{-2N}\right)\;.
\end{equation}
This is nothing else than a rational function approximation to the original function $f(z)$ since $P_i$ and $Q_i$ are polynomials of the designated order. The associated solution for the spectral density reads
\begin{equation}\label{sol}
    \sigma_N(t)=\sum_{i=1}^N B_i^N\delta(t-t_i)\,,\qquad \text{with $B_i^N$ computable in terms of the Pad\'{e} approximants}\;,
\end{equation}
i.e.~the spectral function is approximated by a (finite) series of $\delta$-functions. This sounds well-suited for particle physics where peaks in the spectral functions correspond to particles\footnote{$\delta$-peaks correspond to stable particles, while finite peaks with a certain width to unstable particles. An inherent shortcoming of Pad\'{e} approximation is thus that unstable particles are also replaced by $\delta$-peaks.}. The mass estimates, the $t_i$'s, precisely correspond to the poles of the Pad\'{e} approximant. We will not need the $B_i^N$'s, but they are pivotal to answer another important question: when does a set of moments correspond to a positive spectral function, i.e.~a probability distribution? The answer is when the residues, i.e.~the $B_i^N$'s, are positive numbers. We refer to \cite{moment1,moment2,moment3,moment4} for details and proofs, including the interesting properties of the Pad\'{e} approximants. We limit ourselves to mention here the link with orthogonal polynomials and the location of the poles w.r.t.~the cut structure of $f(z)$: the $Q_{N}$ form a set of polynomials orthogonal over $[0,s_0]$ with weight $\sigma(s)$. Consequently, their zeroes $z_*$ will all be real, different and lying in the interval $]0,s_0[$. Essentially, the poles of the rational approximant have replaced the branch cut of the original function. Now it is clear that any ``reasonable approximative calculation scheme'' (being the moment problem) capable of generating poles in the subtracted result $\mathcal{F}^{\text{sub}}(p^2)$, shall also give poles for $\mathcal{F}(p^2)$, since the difference between both expression is just a polynomial in $p^2$ with infinite coefficients. Thus, we could equally well search for the poles of $f(z)$ in terms of $z (\equiv -1/p^2)$, which we know to be located in the interval $]0,s_0[$, viz.~for $p^2\in ]-\infty,-1/s_0[$. An expansion at small $p^2$ corresponds to a Laurent expansion in $\frac{1}{z}$ near $z\sim\infty$. By power counting, we see that $f(z)\sim \frac{1}{z}$ for $z\sim \infty$, consistent with a $[N,N-1]$ Pad\'{e} approximant. At lowest order, the Pad\'{e} approximant will in general look like
\begin{equation}\label{pad1}
 \frac{P_0(z)}{Q_1(z)}= \frac{-\nu_0'^2}{\nu'_1-\nu_0'z}\;,
\end{equation}
which in return will lead to a mass estimate
\begin{equation}\label{pad2}
  m_{glueball}=\sqrt{\frac{\nu_0'}{\nu_1'}}\;.
\end{equation}
Returning to the question of how we are actually approximating the original (one-loop) correlation function: we start from a (necessarily subtracted) one-loop approximation to the composite operator's propagator using a nonperturbative input gluon propagator. Since in any glueball channel, it is, to our knowledge, quasi impossible to resum classes of diagrams using approximations as in NJL-models for quark bound states, it looks unfeasible to explicitly construct a pole in this correlator by resummation techniques\footnote{This does not exclude to solve, also rather approximately, the Bethe-Salpeter bound state equation as in \cite{Strauss:2012kqa,Meyers:2012ka} to find a mass estimate.}. Since we anyhow have at our disposal a first order approximation for the bound state correlation function, the proposed Pad\'{e} approximation amounts to replace this first order function with a rational one that (i) does have a pole, and (ii) approximates the original function well. From the latter perspective, we point out that for $N\to\infty$, the corresponding Pad\'{e} approximant will converge to the original (one-loop) function $f(z)$ everywhere except on the branch cut of $f(z)$ where the poles of the approximant will pile up, with the smallest pole moving closer to the branch point of $f(z)$ \cite{moment3}. Though, since the function $f(z)$ that we are Pad\'{e} approximating is only a lowest order truncation of the full correlator, it would seem obsolete to consider a high order Pad\'{e} approximant. To merely illustrate how ``well'' a lowest order Pad\'{e} approximant performs, we have displayed in Figure 1 the function $f(z)$ defined in eq.~\eqref{b8c} together with its approximation \eqref{pad1} for the $SU(3)$, $0^{++}$ case. We recall that the approximation is in terms of large $z$ and we see that the original function $f(z)$ ---not displaying a pole--- is quite well replaced by a rational function ---with pole. The function $f(z)$ does not exist for $0\leq z \leq s_0$ as this corresponds to the original branch cut in momentum squared space.
\begin{figure}[t]
  \begin{center}
    \includegraphics[width=14cm]{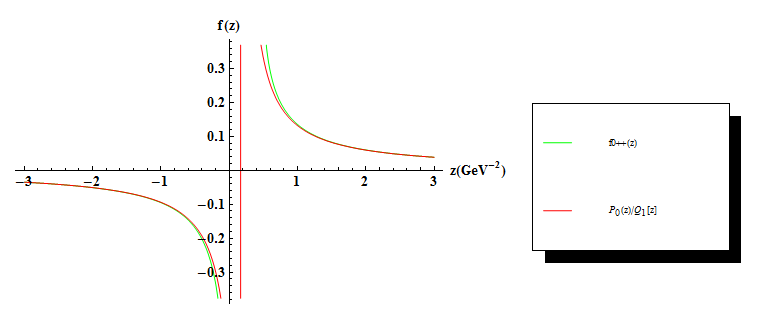}
  \end{center}
  \caption{Comparison of the function $f(z)$ and its lowest order Pad\'{e} approximant.}
\end{figure}
With the hindsight that Pad\'{e} approximation does solve the Hausdorff moment problem, a lowest order Pad\'{e} approximation does correspond to keeping only the first two moments of the moment problem. One might wonder why it was a priori necessary to go through all the troubles of deriving the (one-loop) spectral representation of the correlation function, why not rather compute for Euclidean momenta $p^2\geq0$ the (subtracted) one loop Feynman integral and directly\footnote{One could even think of first expanding and then performing the loop integral to further simply the computations at hand.} expand this up to order $p^2$? This recipe would also immediately be applicable to whatever gluon propagator one would like to use as input. Though, we believe this is the wrong way to proceed, since in that case also unphysical pieces of the, in some approximation computed, correlation function will sneak into the Pad\'{e} approximation, e.g.~from using a gluon propagator with cuts and/or poles in the complex plane as with the RGZ one or as with the fitting form proposed in \cite{Aguilar:2011ux}. With the current setup, we are assured that only the physical piece (positive spectral density, branch cut along the negative real axis) enter the rational approximation scheme.

\subsection{Applications}
\subsubsection{(Pseudo)scalar and (pseudo)tensor glueball operator in $d=4$}
Having armed ourselves with the necessary technology, we are ready to present some results. First, we consider the first 2 moments corresponding to the spectral densities \eqref{a01}-\eqref{a1}. Figure 2(a) shows the mass estimate in terms of the variable subtraction scale $T$ for the gauge group $SU(3)$, based on the RGZ fit \eqref{Dg2},\eqref{a-1}. A simple power counting argument reveals that the (pseudo)scalar spectral representation is well-defined (finite) after 3 subtractions, while the (pseudo)tensor sector needs 7 subtractions.
\begin{figure}[t]
  \begin{center}
    \subfigure[Estimates in terms of the subtraction scale.]{\includegraphics[width=7cm]{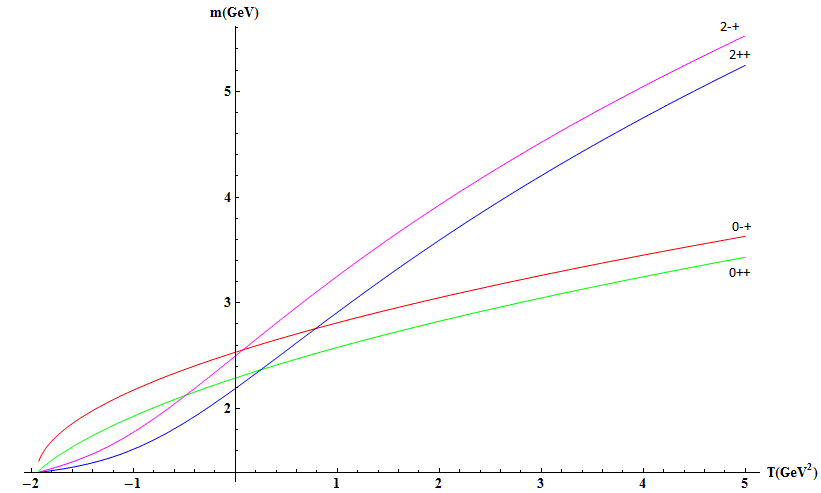} \label{fig1a}}
    \hspace{1cm}
    \subfigure[Averaged estimates in terms of the Gaussian width.]{\includegraphics[width=7cm]{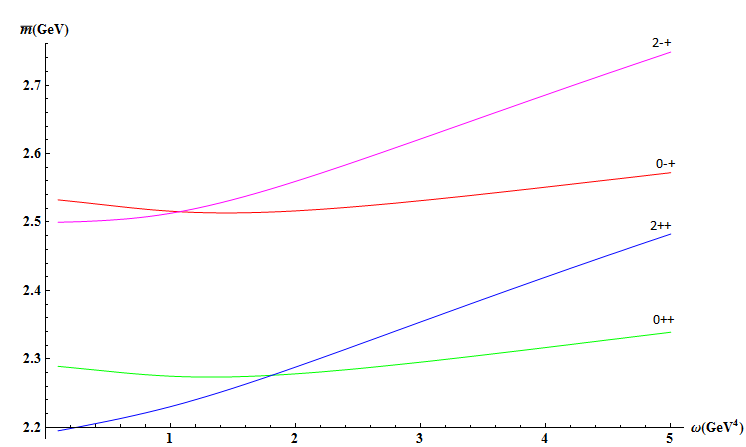}\label{fig1b}}
  \end{center}
  \caption{Glueball masses for $d=4~SU(3)$.}
\end{figure}
\begin{table}[t]
\centering
\begin{tabular}{|c|c|c|c|c|c|}
  \hline
  $J^{PC}$ & this work & \cite[Table XXI]{Chen:2005mg} & \cite[Table 1]{Teper:1998kw} & \cite[Table 14]{Lucini:2004my} & \cite[Table 1]{Szczepaniak:2003mr}\\
  \hline
  $0^{++}$ & 2.27~\text{GeV} & 1.71~\text{GeV} & 1.60~\text{GeV} &1.56~\text{GeV}&1.98~\text{GeV}\\
    $2^{++}$ & 2.34~\text{GeV}  &  2.39~\text{GeV}& 2.27~\text{GeV}&2.10~\text{GeV}&2.42~\text{GeV}\\
  $0^{-+}$ & 2.51~\text{GeV} &   2.56~\text{GeV}& 2.18~\text{GeV}&no data&2.22~\text{GeV}\\
  $2^{-+}$ & 2.64~\text{GeV} &3.04~\text{GeV}&3.10~\text{GeV}&no data&3.09~\text{GeV}\\
  \hline
\end{tabular}
\caption{Optimal mass estimates of this work compared to other approaches for $d= 4~SU(3)$. We have listed the central values, for the corresponding errors we refer to the original works.}
\end{table}
As tried in \cite{Narison:1996fm,Dudal:2010cd}, we could search for an optimal $T$ in terms of which there is a minimal dependence of physical variables on the a priori free $T$, following the spirit of \cite{Stevenson:1981vj}. Unfortunately, there is no optimal $T$ to be found. In fact, for all allowed values of $T>-\tau_0$ there exists a mass estimate. Therefore, we suggest to sample over all possible $T$'s to get an average mass estimate. A logical choice seems to call for a Gaussian distribution centered at zero momentum subtraction (the latter being a rather common choice), whilst allowing for a variable width $\omega>0$. We will then fix $\omega$ using the principle of minimal sensitivity (PMS). Thus,
\begin{equation}\label{distr}
\overline m(\omega)= \frac{\int_{-\tau_2}^\infty \d T m(T) e^{-\frac{T^2}{\omega}}}{\int_{-\tau_2}^\infty \d T e^{-\frac{T^2}{\omega}}}\;.
\end{equation}
It is clear from Figure 2(b) that there is a minimal $\omega$-dependence in the scalar and pseudoscalar case. Albeit less clear from that same Figure 2(b), the (pseudo)tensor case displays an inflection point. After numerically computing the optimal values, $\omega_{PMS}(0^{++},0^{-+},2^{++},2^{-+})\approx(1.32, 1.45, 2.82, 3.32)~\text{GeV}^4$, the optimal mass averaged estimates are shown in Table 1, along some lattice and a Hamiltonian quasi-particle model values, taken from the quoted  papers. Where necessary, we converted to GeV units by using the typical value $\sqrt{\sigma}=0.44$ GeV, notice that this is essentially the value for the string tension discussed in \cite{Sommer:1993ce,Bali:1992ru}.  We simply picked the central value of the lattice papers to get an idea of their mass estimates. In order to get a rudimentary error estimate on the reported mean value $\overline m$, we computed the standard deviation $\overline\sigma$ associated to the distribution \eqref{distr} for $\omega=\omega_{PMS}$, finding $\overline\sigma(0^{++},0^{-+},2^{++},2^{-+})\approx(0.26, 0.27, 0.67, 0.79)~\text{GeV}$.

Similarly, we may consider the $SU(2)$ cases based on the expressions \eqref{Dg2} and \eqref{a-1b}, leading to Figures 3(a),3(b) and the mass estimates shown in Table 2. For the record, the optimal widths are provided by $\omega_{PMS}(0^{++},0^{-+},2^{++},2^{-+})\approx(1.37, 1.54, 2.95, 3.36)~\text{GeV}^4$. The standard deviations in this case are given by $\overline\sigma(0^{++},0^{-+},2^{++},2^{-+})\approx(0.25, 0.27, 0.66 , 0.72 )~\text{GeV}$. The degeneracy between the pseudoscalar and pseudotensor is only apparent due to rounding errors.  One may notice that there is only a little difference between the $SU(2)$ and $SU(3)$ mass estimates in the current approximation, this should not come as a surprise given that the $d=4$ lattice data for $SU(2)$ and $SU(3)$ are closely resemblant see e.g.~\cite{Cucchieri:2007zm}.
\begin{figure}[t]
  \begin{center}
    \subfigure[Estimates in terms of the subtraction scale.]{\includegraphics[width=7cm]{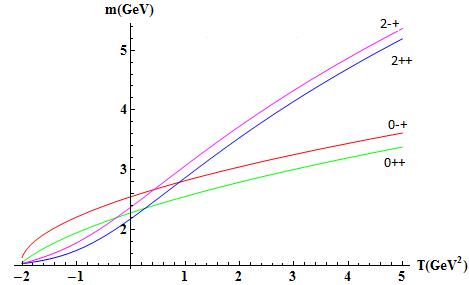} \label{fig1a}}
    \hspace{1cm}
    \subfigure[Averaged estimates in terms of the Gaussian width.]{\includegraphics[width=7cm]{massabarSU3.png}\label{fig1b}}
  \end{center}
  \caption{Glueball masses for $d=4~SU(2)$. We have listed the central values, for the corresponding errors we refer to the original works.}
\end{figure}
\begin{table}[t]
\centering
\begin{tabular}{|c|c|c|c|}
  \hline
  $J^{PC}$ & this work & \cite[Table 1]{Teper:1998kw} & \cite[Table 14]{Lucini:2004my} \\
  \hline
  $0^{++}$ & 2.26~\text{GeV} & 1.65~\text{GeV} & 1.66~\text{GeV} \\
    $2^{++}$ & 2.33~\text{GeV}  &  2.47~\text{GeV}& 2.44~\text{GeV}\\
  $0^{-+}$ & 2.53~\text{GeV} &   2.87~\text{GeV}& no data\\
  $2^{-+}$ & 2.53~\text{GeV} &3.28~\text{GeV}&no data\\
  \hline
\end{tabular}
\caption{Optimal mass estimates of this work compared to other approaches for $d=4~SU(2)$. We have listed the central values, for the corresponding errors we refer to the original works. }
\end{table}
\subsubsection{Scalar and tensor glueball operator in $d=3$}
As a final application, we consider 2 glueball states in $d=3$, based on eq.~\eqref{Dg2} and eqns.~\eqref{b01}-\eqref{b02}. In $d=3$, we need one less subtraction compared to $d=4$, thus $r=2$, resp.~$r=6$ for the scalar, resp.~tensor glueball. The quoted analytical work \cite{Simonov:2006re} relies on the field correlator method, while \cite{Leigh:2006vg} on the gauge invariant Karabali \& Nair variables \cite{Karabali:1995ps,Karabali:1996je}. The corresponding numbers can be found in Table 3. Unfortunately, it is clear from Figures 4(a), 4(b) we can neither extract an optimal width $\omega$ nor an optimal subtraction scale $T$ in the $d=3$ case. At best, we can put an estimate  2-3.5~GeV with the $2^{++}$ probably heavier given its larger mass over most of the shown $\omega$- (or $T$-) interval.
\begin{figure}[h]
  \begin{center}
    \subfigure[Estimates in terms of the subtraction scale.]{\includegraphics[width=7cm]{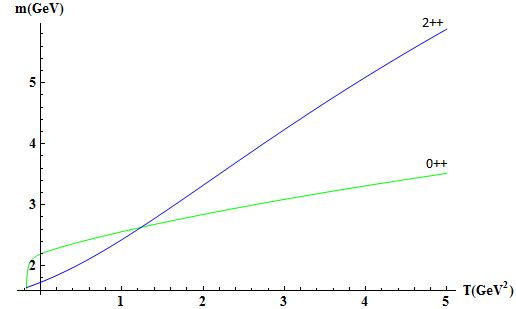} \label{fig1a}}
    \hspace{1cm}
    \subfigure[Averaged estimates in terms of the Gaussian width.]{\includegraphics[width=7.5cm]{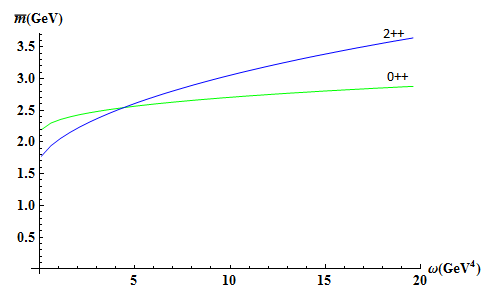}\label{fig1b}}
  \end{center}
  \caption{Glueball masses for $d=3~SU(2)$.}
\end{figure}
\begin{table}[h]
\centering
\begin{tabular}{|c|c|c|c|c|}
  \hline
  $J^{PC}$  & \cite[Table 1]{Johnson:2000qz} & \cite[Table 1]{Simonov:2006re} & \cite[Table 1]{Leigh:2006vg}\\
  \hline
  $0^{++}$ &  2.07~\text{GeV} & 2.02~\text{GeV} & 2.02~\text{GeV} \\
    $2^{++}$ &  3.44~\text{GeV}& 2.87~\text{GeV}&no result\\
  \hline
\end{tabular}
\caption{Mass estimates for $d=3~SU(2)$. }
\end{table}

\section{Discussion}
We have discussed, with a recently introduced IR moment technique and ensuing rational (Pad\'{e}) approximation, mass estimates for a set of glueball states in gluodynamics. The cases $SU(3)$ and $SU(2)$ have been reported in $d=4$, while in $d=3$ the case of $SU(2)$ has been considered.

Our calculations have relied on the tree level confining gluon RGZ propagators, which encode nonperturbative information on the Gribov horizon and on dimension two condensates.  Even if we have been working in a lowest order approximation, the fact that in most cases the Pad\'{e} estimates for the masses reside in the same ballpark as other analytical and/or lattice estimates ---which themselves sometimes show a widespread range of numbers--- gives credit to our infrared moments method in general and to the expectation that important nonperturbative information is already  present in the gluon propagator. Future work should be directed towards adding higher loop contributions, e.g.~in the Refined Gribov-Zwanziger framework, to the glueball correlators and see how they influence the mass estimates, although it seems  not easy  to obtain the higher order corrections to the spectral densities in closed form. In such instance, the Pad\'{e} method might also be of help since the first few moments of a specific correlator could be computed directly from the momentum space expression of the correlator and, as mentioned in the main text  below eq.~\eqref{sol}, the rational approximants can be used to decide whether these moments can belong to a positive spectral density or not \cite{moment1}. It would also be instructive to investigate how sensitive the Pad\'{e} results are to different numerical/functional prescriptions for gluon propagators, as obtained, for instance,  in \cite{Strauss:2012dg,Aguilar:2011ux,Aguilar:2004sw,Aguilar:2008xm,Fischer:2008uz,Tissier:2010ts,Boucaud:2011ug}.

\section*{Acknowledgments}
We thank L.~F.~Palhares for a careful reading of the manuscript. The Conselho Nacional de Desenvolvimento Cient\'{\i}fico e
Tecnol\'{o}gico (CNPq-Brazil), the Faperj, Funda{\c{c}}{\~{a}}o de
Amparo {\`{a}} Pesquisa do Estado do Rio de Janeiro, the Latin
American Center for Physics (CLAF), the SR2-UERJ,  the
Coordena{\c{c}}{\~{a}}o de Aperfei{\c{c}}oamento de Pessoal de
N{\'{\i}}vel Superior (CAPES) and the Research-Foundation Flanders (via the Odysseus grant of F.~Verstraete) are gratefully acknowledged.

\end{document}